\documentclass{ws-ijmpa}
\usepackage[super,compress]{cite}
\usepackage{graphicx}
\usepackage{mathtools}
\usepackage{txfonts}
\UseRawInputEncoding
\begin{document}
\markboth{G. L. Klimchitskaya, C. C. Korikov \& V. M. Mostepanenko}{Nonequilibrium Casimir Pressure Between Graphene-Coated Plates}

%
\catchline{}{}{}{}{}
%

\title{\uppercase{Nonequilibrium Casimir pressure for two graphene-coated plates:
quantum field theoretical approach}}

\author{\uppercase{G. L. Klimchitskaya,${}^{1,2}$ C. C. Korikov${}^3$} \lowercase{and}
\uppercase{V. M. Mostepanenko${}^{1,2}$}}
\address{${}^1$Central Astronomical Observatory at Pulkovo of the
Russian Academy of Sciences, Saint Petersburg,
196140, Russia\\
${}^2$Peter the Great Saint Petersburg
Polytechnic University, Saint Petersburg, 195251, Russia\\
${}^3$Huawei Noah's Ark Lab, Krylatskaya str. 17, Moscow 121614, Russia \\
g.klimchitskaya@gmail.com, constantine.korikov@gmail.com,  vmostepa@gmail.com
}

\maketitle


\begin{abstract}
We consider the nonequilibrium Casimir pressure in the system of two
parallel graphene-coated plates one of which is either warmer or
cooler than the environment. The electromagnetic response of
graphene coating characterized by the nonzero energy gap and chemical
potential is described in the framework of the Dirac model by means
of the polarization tensor. It is shown that the magnitude of the
nonequilibrium Casimir pressure on a warmer plate than the environment
is larger and on a cooler plate is smaller than the magnitude of the
standard Casimir pressure in the state of thermal equilibrium. According
to our results, the spatially local theory underestimates the role of
the effects of nonequilibrium. This underestimation increases for a
smaller chemical potential of the graphene coating and at lower
temperatures of the cooled plate. Possible applications of the obtained
results are indicated.
\end{abstract}

\ccode{PACS numbers: 12.20.-m, 12.20.Ds, 68.65.Pq, 05.70.Ln}

\newcommand{\ve}{{\varepsilon}}
\newcommand{\ix}{{(i\xi_{E,l},k,\Delta,\mu,T_P)}}
\newcommand{\up}{{(u,k,\mu,T_p)}}
\newcommand{\qe}{{q_{E,l}(k)}}
\newcommand{\eq}{{q_{E,l}^{\,\varepsilon}(k)}}
\newcommand{\tq}{{\tilde{q}_{E,l}(k)}}
\newcommand{\feq}{{F_{\rm eq}^{\rm SiO_2}}}
\newcommand{\nfeq}{{F_{\rm neq}^{\rm SiO_2}}}
\newcommand{\po}{{(\omega,k,T_p)}}
\newcommand{\wq}{{q^{\,\varepsilon}}}
\newcommand{\Ww}{{(u,\omega,k,\mu,T_p)}}
\newcommand{\Tw}{{\widetilde{w}}}

\section{Introduction}	

It is well known that the Casimir effect, i.e., the attractive force acting
between two parallel plates caused by fluctuations of the electromagnetic
field,\cite{1} is the physical phenomenon which occurs in thermal equilibrium.
What this means is temperatures of both plates are equal to each other and
to that of the environment. In this case, the unified theory of the
van der Waals and Casimir forces was created by E. M. Lifshitz.\cite{2}

The nonequilibrium version of the Casimir effect takes place in situations
when either the temperatures of the plates are not the same or at least one
of them is not equal to the environmental temperature. The generalization of
the Lifshitz theory for situations of this kind was developed under a condition
that each plate is in the state of local thermal equilibrium.\cite{3,4,5,6,6a}
Similar formalism was also developed for the Casimir-Polder interaction of a
polarizable particle with a material plate.\cite{7,8,8a} The obtained
results were generalized for the case of arbitrarily shaped
bodies.\cite{9,10,11,12,13,14,15,16} The possibilities of observing the
nonequilibrium Casimir force were discussed in Refs. \citen{17,18}.

In out-of-thermal-equilibrium conditions, it may happen that not only the
Planckian distribution of the electromagnetic radiation, but the material
properties of the interacting bodies as well, explicitly depend on
temperature as a parameter. It was shown\cite{19} that the developed theory
of the nonequilibrium Casimir interaction\cite{5,6} can be adapted for use
in this important case.

The novel material whose reflectivity ptoperties strongly depend on temperature
is graphene, which is a two-dimensional sheet of carbon atoms packed in the
hexagonal lattice.\cite{20} At energies below approximately 3~eV, the
quasiparticles in graphene are well described by the Dirac equation where the
speed of light $c$ is replaced with the Fermi velocity $v_F$. It was shown that
the thermal regime for the Casimir force between two graphene sheets sets in at
much shorter separations than between metallic plates.\cite{21} This result was
confirmed experimentally.\cite{22,23} That is why the nonequilibrium Casimir
force in graphene systems is the subject matter which has a great potential.

Investigation of the Casimir-Polder interaction between nanoparticles and
graphene sheets, both freestanding in vacuum and deposited on a substrate, was
performed quite recently.\cite{24,25,26,27} The nonequilibrium Casimir-Polder
force behavior was studied as a function of the plate-nanoparticle separation,
energy gap parameter and chemical potential of a graphene sheet, and on the
presence of a material substrate. Furthermore, the Casimir pressure between
two graphene-coated plates out of thermal equilibrium was investigated in the
special case of gapless graphene sheets using the Kubo formalism in the
spatially local approximation.\cite{28}

In this paper, we consider the nonequilibrium Casimir interaction of two
material plates coated with gapped and doped graphene. The electromagnetic
response of graphene is described in the framework of quantum field theory
by means of the polarization
tensor which takes full account for the effects of spatial nonlocality.
It is shown that for a heated plate, as compared to the environment, the
magnitude of the nonequilibrium Casimir pressure is larger than that of the
equilibrium one. For a cooled plate, the situation is in complete contrast.
According to our results, the local theory underestimates the role of the
nonequilibrium effects. The contribution of these effects decreases with
increasing chemical potential. Finally, the contribution of the spatially
nonlocal effects is larger for the cooler graphene-coated plate than the
environment and smaller for the warmer plate.

\section{Electromagnetic Response of Graphene in Terms of Polarization Tensor}

The reflection coefficients from graphene sheet deposited on a plate with the
dielectric permittivity  $\ve\equiv\ve(\omega)$ are expressed via the polarization
tensor of graphene\cite{27,29}
\begin{eqnarray}
&&
R_{\rm TM}\po=\frac{\hbar k^2(\ve q-q^{\ve})+
q\wq\Pi_{00}}{\hbar k^2(\ve q+\wq)+
q\wq\Pi_{00}},
\nonumber \\
&&
R_{\rm TE}\po=\frac{\hbar k^2(q-\wq)-\Pi}{\hbar k^2
(q+\wq)+\Pi},
\label{eq1}
\end{eqnarray}
\noindent
where $q\equiv q(\omega,k)=(k^2-\omega^2/c^2)^{1/2}$,
$\wq\equiv \wq(\omega,k)=(k^2-\ve\omega^2/c^2)^{1/2}$,
$k$ is the magnitude of the wave vector projection on the
plane of graphene, $\Pi_{\beta\gamma}\equiv\Pi_{\beta\gamma}\po$
with $\beta,\,\gamma=0,\,1,\,2$ is the polarization tensor of
graphene, which also depends on the energy gap $\Delta$ and
chemical potential $\mu$, $T_p$ is the temperature of the
graphene-coated plate, and
$\Pi\equiv k^2\Pi_{\beta}^{\,\beta}-q^2\Pi_{00}$.

In the framework of the Dirac model, the explicit expressions
for the polarization tensor of graphene at pure imaginary Matsubara
frequencies $i\xi_l=2\pi ik_BTl/\hbar$, where $k_B$ is the
Boltzmann constant, $l=0,\,1,\,2,\,\ldots\,$, were found
for use in calculations of the Casimir force.\cite{30,31,32}
They were analytically continued to the real frequency axis
for application in calculations of the reflectivity properties
and electrical conductivity of graphene.\cite{27,33,34}
Below we present the expressions for $\Pi_{00}$ and $\Pi$ in
the most convenient explicit form.

First and foremost the quantities $\Pi_{00}$ and $\Pi$ have
different forms in the frequency regions $\omega\geqslant v_Fk$
and $\omega<v_Fk$. We begin with the region $\omega\geqslant v_Fk$.
If $\hbar cp<\Delta$, where
$p\equiv p(\omega,k)=(\omega^2-v_F^2k^2)^{1/2}/c$,
one obtains\cite{27,33}
\begin{equation}
\Pi_{00}=-\frac{2\alpha k^2}{cp^2}\,\Phi_1+
\frac{4\alpha\hbar c^2p}{v_F^2}\int\limits_{\widetilde{D}}^{\infty}
du\,\Tw
\left[1-\frac{1}{2cp}\sum_{\lambda=\pm 1}\lambda B_1(cp u+\lambda\omega)\right],
\label{eq2}
\end{equation}
\noindent
where $\alpha$ is the fine structure constant and
\begin{eqnarray}
&&
\Phi_1\equiv\Phi_1(\omega,k,\Delta)=\Delta-\hbar cp\left[1+\frac{\Delta^2}{(\hbar cp)^2}
\right]\,{\rm arctanh}\frac{\hbar cp}{\Delta},
\nonumber\\
&&
\Tw\equiv\Tw\Ww=\sum_{\kappa=\pm 1}\left(e^{\widetilde{B}u+\kappa\frac{\mu}{k_BT_p}}
+1\right)^{-1}\!\!\!,
\nonumber\\
&&
\widetilde{D}\equiv\widetilde{D}(\omega,k,\Delta)=\frac{\Delta}{\hbar cp}, \qquad
\widetilde{B}\equiv\widetilde{B}(\omega,k,T_p)=\frac{\hbar cp}{2k_BT_p},
\nonumber \\
&&
B_1(x)=\frac{x^2-v_F^2k^2}{(x^2-v_F^2k^2A)^{1/2}}, \qquad
A\equiv A(\omega,k,\Delta)=1-\frac{\Delta^2}{(\hbar cp)^2} .
\label{eq3}
\end{eqnarray}
It is seen that due to the condition $\hbar cp<\Delta$
the quantity $\Pi_{00}$ is real.

Under the same condition, the quantity $\Pi$ is also real.
It is given by
\begin{equation}
\Pi=\frac{2\alpha k^2}{c}\,\Phi_1+
\frac{4\alpha\hbar p\omega^2}{v_F^2}\int\limits_{\widetilde{D}}^{\infty}
du\,\Tw
\left[1-\frac{cp}{2\omega^2}\sum_{\lambda=\pm 1}\lambda
B_2(cp u+\lambda\omega)\right],
\label{eq4}
\end{equation}
\noindent
where
\begin{equation}
B_2(x)=\frac{x^2+v_F^2k^2(1-A)}{(x^2-v_F^2k^2A)^{1/2}}.
\label{eq5}
\end{equation}

Under the opposite condition $\hbar cp\geqslant\Delta$, the quantities
$\Pi_{00}$ and $\Pi$ are complex. Thus, $\Pi_{00}$ takes the
form\cite{27,33}
\begin{eqnarray}
&&
\Pi_{00}=-\frac{2\alpha k^2}{cp^2}\,\Phi_2+
\frac{4\alpha\hbar c^2p}{v_F^2}\left\{\int\limits_{\widetilde{D}}^{u_1}
du\,\Tw
\left[1-\frac{1}{2cp}\sum_{\lambda=\pm 1} B_1(cp u+\lambda\omega)\right]
\right.
\nonumber\\
&&~~~~~~~~~~~~~~~~~
\left.+\int\limits_{{u_1}}^{\infty}
du\,\Tw
\left[1-\frac{1}{2cp}\sum_{\lambda=\pm 1}\lambda B_1(cp u+\lambda\omega)\right]
\right\},
\label{eq6}
\end{eqnarray}
\noindent
where
\begin{eqnarray}
&&
\Phi_2\equiv\Phi_2(\omega,k,\Delta)=\Delta-\hbar cp\left[1+\frac{\Delta^2}{(\hbar cp)^2}
\right]\,\left({\rm arctanh}\frac{\Delta}{\hbar cp}+i\frac{\pi}{2}\right),
\nonumber\\
&&
u_{1,2}=\frac{1}{cp}(\omega\mp v_Fk\sqrt{A}).
\label{eq7}
\end{eqnarray}

The quantity $\Pi$ under the condition $\hbar cp\geqslant\Delta$
is defined as
\begin{eqnarray}
&&
\Pi=\frac{2\alpha k^2}{c}\,\Phi_2+
\frac{4\alpha\hbar p\omega^2}{v_F^2}\left\{\int\limits_{\widetilde{D}}^{u_1}
du\,\Tw
\left[1-\frac{cp}{2\omega^2}\sum_{\lambda=\pm 1}
B_2(cp u+\lambda\omega)\right]
\right.
\nonumber\\
&&~~~~~~~~~~~~~~~~~
\left.+\int\limits_{u_1}^{\infty}
du\,\Tw
\left[1-\frac{cp}{2\omega^2}\sum_{\lambda=\pm 1}\lambda
B_2(cp u+\lambda\omega)\right]
\right\}.
\label{eq8}
\end{eqnarray}

Next, we consider the region $\omega<v_Fk$. Here, for $\Pi_{00}$
one obtains\cite{27,33}
\begin{equation}
\Pi_{00}=\frac{\alpha\hbar k^2}{\tilde{q}}\Psi(D)+
\frac{4\alpha\hbar c^2\tilde{q}}{v_F^2}\int\limits_{D}^{\infty}\!\!du\,w
\left[1-\frac{1}{2}\sum_{\lambda=\pm 1}\!\!\lambda
\frac{1- u^2+2\lambda\widetilde{\gamma} u}{\left(1-u^2+
2\lambda\widetilde{\gamma} u+
D^2+\widetilde{\gamma}^2 D^2\right)^{1/2}}\right],
\label{eq9}
\end{equation}
\noindent
where $\Psi(x)=2[x+(1-x^2)\,{\rm arctan}(1/x)]$ and
\begin{eqnarray}
&&
\tilde{q}\equiv\tilde{q}(\omega,k)=\frac{1}{c}\sqrt{v_F^2k^2-\omega^2}, \qquad
\widetilde{\gamma}=\widetilde{\gamma}(\omega,k)
=\frac{\omega}{\sqrt{v_F^2k^2-\omega^2}},
\nonumber\\
&&
D\equiv D(\omega,k,\Delta)=\frac{\Delta}{\hbar c\tilde{q}},\qquad
B\equiv B(\omega,k,T_p)=\frac{\hbar c \tilde{q}}{2k_BT_p}
\nonumber\\
&&
w\equiv w\Ww=\sum_{\kappa=\pm 1}\left(e^{Bu+\kappa\frac{\mu}{k_BT_p}}
+1\right)^{-1}\!\!\!.
\label{eq10}
\end{eqnarray}

Finally, for the quantity $\Pi$ in the region $\omega<v_Fk$ it holds
\begin{equation}
\Pi={\alpha\hbar k^2}{\tilde{q}}\Psi(D)+
\frac{4\alpha\hbar \omega^2\tilde{q}}{v_F^2}\int\limits_{D}^{\infty}\!\!du\,w
\left[1-\frac{1}{2}\sum_{\lambda=\pm 1}\!\!\lambda
\frac{(1-\lambda\widetilde{\gamma}^{-1} u)^2-
(\widetilde{\gamma}^{-2}+1)D^2}{\left(1-u^2+
2\lambda\widetilde{\gamma} u+
D^2+\widetilde{\gamma}^2 D^2\right)^{1/2}}\right].
\label{eq11}
\end{equation}
\noindent
Thus, the reflection coefficients (\ref{eq1}) are defined for all
values of $\omega$ and $k$.

\section{Nonequilibrium Casimir Pressure}
\newcommand{\rf}{R_{\alpha}(v,t,T_p^{(1)})}
\newcommand{\rs}{R_{\alpha}(v,t,T_p^{(2)})}

For the sake of definiteness, we consider the configuration
of two parallel plates at a distance $a$ made of the same
material with the temperature-independent dielectric permittivity
$\ve(\omega)$ coated by a sheet of gapped and doped graphene.
We assume that the first graphene-coated plate is kept at
the environmental temperature $T_p^{(1)}=T_E=300~$K,
whereas the temperature of the second graphene-coated plate
$T_p^{(2)}$ can be either lower or higher that $T_E$.

The nonequilibrium Casimir pressure on the second plate is
given by\cite{6,6a,19}
\begin{equation}
P_{\rm neq}(a,T_p^{(1)},T_p^{(2)})=\widetilde{P}_{\rm eq}(a,T_p^{(1)},T_p^{(2)})
+\Delta P_{\rm neq}(a,T_p^{(1)},T_p^{(2)}),
\label{eq12}
\end{equation}

Here, $\widetilde{P}_{\rm eq}$ is the quasi equilibrium contribution
given by the linear combination of two Lifshitz-type formulas\cite{2,19,35}
\begin{eqnarray}
&&
\widetilde{P}_{\rm eq}(a,T_p^{(1)},T_p^{(2)})=-\frac{k_B}{2\pi}\left[
T_p^{(1)}\sum_{l=0}^{\infty}{\vphantom{\sum}}^{\prime}
\int\limits_{0}^{\infty}q_l^{(1)}kdk\right.
\nonumber\\
&&~~~~~~~~~~~~~~~~~~~~~~~~\times
\sum_{\alpha}\frac{R_{\alpha}(i\xi_l^{(1)}\!\!,k,T_p^{(1)})
R_{\alpha}(i\xi_l^{(1)}\!\!,k,T_p^{(2)})}{e^{2q_l^{(1)}a}-
R_{\alpha}(i\xi_l^{(1)}\!\!,k,T_p^{(1)})
R_{\alpha}(i\xi_l^{(1)}\!\!,k,T_p^{(2)})}
\nonumber\\
&&~~\left.+
T_p^{(2)}\sum_{l=0}^{\infty}{\vphantom{\sum}}^{\prime}\!
\int\limits_{0}^{\infty}q_l^{(2)}kdk\sum_{\alpha}
\frac{R_{\alpha}(i\xi_l^{(2)}\!\!,k,T_p^{(1)})
R_{\alpha}(i\xi_l^{(2)}\!\!,k,T_p^{(2)})}{e^{2q_l^{(2)}a}-
R_{\alpha}(i\xi_l^{(2)}\!\!,k,T_p^{(1)})
R_{\alpha}(i\xi_l^{(2)}\!\!,k,T_p^{(2)})}\right].
\label{eq16}
\end{eqnarray}
\noindent
The quantities $q_l^{(j)}$ and $R_{\alpha}(i\xi_l^{(j)}\!\!,k,T_p^{(f)})$ are
obtained from the definition of $q(\omega,k)$ and the reflection coefficients
(\ref{eq1}) by putting $\omega=i\xi_l^{(j)}$, where $\xi_l^{(j)}$ are the
Matsubara frequencies calculated at the temperatures of the first and
second plates. The quantities $\Pi_{00}$ and $\Pi$ taken at the pure
imaginary Matsubara frequencies entering (\ref{eq16}) are connected with
(\ref{eq9}) and (\ref{eq11}) by the same replacement. In doing so, one
should pay special attention to the correct choice of the branch of the
square roots (see the expressions in Ref.~\refcite{36}).

If the dielectric response of both plates does not depend on $T$
(this is the case for SiO${}_2$ plates with no graphene coating), the
reflection coefficients in (\ref{eq16}) depend on $T$ only through
the Matsubara frequencies. In this case (\ref{eq16}) take the form
\begin{equation}
\widetilde{P}_{\rm eq}(a,T_p^{(1)},T_p^{(2)})=\frac{1}{2}\left[
P_{\rm eq}(a,T_p^{(1)})+P_{\rm eq}(a,T_p^{(2)})\right],
\label{eq13}
\end{equation}
\noindent
where $P_{\rm eq}(a,T_p^{(j)})$ is given by the Lifshitz formula
for two parallel plates kept either at the temperature $T_p^{(1)}$
or $T_p^{(2)}$ with corresponding temperatures of the environment.

Note that when the temperatures of both plates are equal to each
other and to the environment temperature, $T_p^{(1)}=T_p^{(2)}=T_E$,
the pressure (\ref{eq16}) coincides with the standard equilibrium
Casimir pressure given by the Lifshitz formula
\begin{equation}
\widetilde{P}_{\rm eq}(a,T_p^{(1)},T_p^{(1)})=
P_{\rm eq}(a,T_E).
\label{eq17}
\end{equation}
\noindent
In this case, the two contributions on the r.h.s. of (\ref{eq16})
coincide.

The proper nonequilibrium term in (\ref{eq12}) is expressed as \cite{6,6a,19}
\begin{equation}
\Delta P_{\rm neq}(a,T_p^{(1)},T_p^{(2)})=\frac{\hbar c}{64\pi^2a^4}
\int\limits_{0}^{\infty}v^3dv[n(v,T_p^{(1)})-n(v,T_p^{(2)})]
\nonumber
\end{equation}
\vspace*{-3mm}
\begin{equation}
\times \sum_{\alpha}\left[\int\limits_{0}^{1}t\sqrt{1-t^2}dt
\frac{|\rs|^2-|\rf|^2}{|D_{\alpha}(v,t,T_p^{(1)},T_p^{(2)})|^2}\right.
-2\int\limits_{1}^{\infty}t\sqrt{t^2-1}e^{-v\sqrt{t^2-1}}dt
\nonumber
\end{equation}
\begin{equation}
\left.
\times
\frac{{\rm Im}\rf{\rm Re}\rs-
{\rm Re}\rf{\rm Im}\rs}{|D_{\alpha}(v,t,T_p^{(1)},T_p^{(2)})|^2}\right],
\label{eq14}
\end{equation}
\noindent
where the reflection coefficients and the polarization tensor are
presented in (\ref{eq1}), (\ref{eq2}), (\ref{eq4}), (\ref{eq6}),
(\ref{eq8}), (\ref{eq9}), and (\ref{eq11}), and the following notations are introduced:
\begin{eqnarray}
&&
n(v,T_p^{(j)})=\left[\exp\left(\frac{\hbar cv}{2ak_BT_p^{(j)}}
\right)-1\right]^{-1}\!\!\!\!,
\quad v=\frac{2a\omega}{c}, \quad t=\frac{ck}{\omega},
\nonumber\\
&&
D_{\alpha}(v,t,T_p^{(1)},T_p^{(2)})=1-\rf\rs\,e^{iv\sqrt{1-t^2}}.
\label{eq15}
\end{eqnarray}

\section{Computational Results for the Nonequilibrium Casimir
Pressure}

Here, we calculate the nonequilibrium Casimir pressure on the second
plate in situations when its temperature is either lower or higher than
that of the first plate and of the environment. In doing so,
special attention is paid to the role of spatial nonlocality in the
electromagnetic response of graphene described by the polarization
tensor which was not investigated so far.

As was specified in Sec.~3, $T_p^{(1)}=T_E=300~$K. Below we put
the temperature of the second plate equal to either $T_p^{(2)}=500~$K
(higher than $T_E$) or $T_p^{(2)}=77~$K (lower than $T_E$).
In all computations, we consider the graphene sheet deposited
on a SiO${}_2$ substrate as in the recently performed
experiments\cite{22,23} and use the value of the energy gap
$\Delta=0.3~$eV close to that measured in these experiments.

Computations of the nonequilibrium Casimir pressure $P_{\rm neq}$
were performed by Eqs.~(\ref{eq12}), (\ref{eq16}), and (\ref{eq14}). In so doing,
the analytic expressions for the polarization tensor presented
in Sec.~2 have been used. The frequency-dependent
dielectric permittivity of the
SiO${}_2$ plate was obtained based on the tabulated optical data for
its complex index of refraction.\cite{37}

It is convenient to present the computational results in terms of
the three quantities. The first of them is defined as
\begin{equation}
\delta P_{\rm neq}(a,T_p^{(1)},T_p^{(2)})=
\frac{P_{\rm neq}(a,T_p^{(1)},T_p^{(2)})-P_{\rm eq}(a,T_E)}{P_{\rm eq}(a,T_E)},
\label{eq18}
\end{equation}
\noindent
where $P_{\rm neq}$ is given by (\ref{eq12}) and $P_{\rm eq}$ by (\ref{eq17}).
Thus, the quantity (\ref{eq18}) characterizes the relative contribution of
the effects of nonequilibrium to the total nonequilibrium Casimir pressure
$P_{\rm neq}$.

The second relevant quantity is
\begin{equation}
\delta P_{\rm neq}^l(a,T_p^{(1)},T_p^{(2)})=
\frac{P_{\rm neq}^l(a,T_p^{(1)},T_p^{(2)})-
P_{\rm eq}^l(a,T_E)}{P_{\rm eq}^l(a,T_E)},
\label{eq19}
\end{equation}
\noindent
The quantity (\ref{eq19}) is similar to (\ref{eq18}) but calculated with the
spatially nonlocal effects disregarded. This is reached by putting $k=0$ in
expressions (\ref{eq2}), (\ref{eq4}), (\ref{eq6}), and (\ref{eq8}) for
the polarization tensor when performing numerical computations. Therefore
(\ref{eq19}) represents the relative contribution of the effects of
nonequilibrium in the spatially local theory.

Finally, the third quantity calculated below
\begin{equation}
\delta P_{\rm neq}^{(1)}(a,T_p^{(1)},T_p^{(2)})=
\frac{\widetilde{P}_{\rm eq}(a,T_p^{(1)},T_p^{(2)})-
P_{\rm eq}(a,T_E)}{P_{\rm eq}(a,T_E)},
\label{eq20}
\end{equation}
\noindent
where $\widetilde{P}_{\rm eq}$ is given by (\ref{eq13}) and
${P}_{\rm eq}$ by (\ref{eq17}), describes the relative role of the
effects of nonequilibrium in the quasi-equilibrium contribution to
the total pressure defined in (\ref{eq16}).

In Fig.~1 (left), the computational results for the quantities $\delta P_{\rm neq}$,
$\delta P_{\rm neq}^l$, and $\delta P_{\rm neq}^{(1)}$ are shown as the functions
of separation between the graphene-coated plates by the solid, long-dashed, and
short-dashed lines, respectively. The positive values of  $\delta P_{\rm neq}$,
$\delta P_{\rm neq}^l$, and $\delta P_{\rm neq}^{(1)}$ were obtained for the
heated second plate, $T_p^{(2)}=500~$K, whereas for the cooled second plate,
$T_p^{(2)}=77~$K, all these quantities take the negative values.
In Fig.~1 (left) computations were performed for a graphene coating with $\mu=0$.
\begin{figure}[b]
\vspace*{-10.cm}
\centerline{\includegraphics[width=17.5cm]{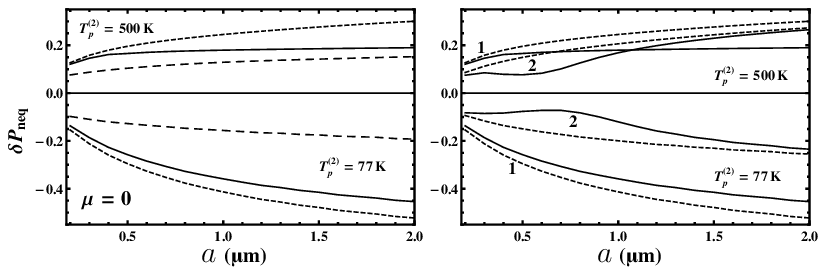}}
\vspace*{-11.3cm}
\caption{The computational results for the relative contributions of the
effect of nonequilibrium to the Casimir pressure, $\delta P_{\rm neq}$,
$\delta P_{\rm neq}^l$, and $\delta P_{\rm neq}^{(1)}$, are shown as the functions
of separation by the solid, long-dashed, and short-dashed lines, respectively.
The graphene coating has (left) $\Delta=0.3$~eV, $\mu=0$ and (right)
$\Delta=0.3$~eV, $\mu=0$ (the pairs of solid and short-dashed lines labeled 1)
and $\Delta=0.3$~eV, $\mu=0.25$~eV (the pairs of solid and short-dashed
lines labeled 2). The lines in the upper half-plane are for $T_p^{(2)}=500$~K
and in the lower half-plane $T_p^{(2)}=77$~K.
 \label{f1}}
\end{figure}

As is seen in Fig.~1 (left), for the values of parameters used in computations, the
magnitude of the nonequilibrium Casimir pressure is larger than the magnitude
of the equilibrium one for a heated plate and smaller than that for a cooled plate.
From a comparison of the solid and long-dashed lines, one can conclude that
the magnitudes of relative contribution of the effects of nonequilibrium
computed with account of spatial nonlocality exceed those computed using the
spatially local theory. Thus, the local theory underestimates the role of
nonequilibrium effects. This underestimation becomes larger for a cooled plate
with increasing separation between the plates. Finally, from the comparison
of the solid and short-dashed lines, it follows that contributions
of the quasi-equilibrium and the proper nonequilibrium terms
in (\ref{eq12}) to the relative role of the effects of nonequilibrium are
of the opposite sign.

In Fig.~1 (right), we illustrate an impact of the chemical potential
of the graphene coating on the relative role of the effects of
nonequilibrium in the Casimir pressure. For this purpose, two pairs
of the solid and short-dashed lines plotted for $\mu=0$ at $T_p^{(2)}=500$~K
and $T_p^{(2)}=77$~K are reproduced from Fig. 1 (left). Both of them are
labeled 1. In addition, by the two pairs of the solid and short-dashed
lines labeled 2, we plot the quantities $\delta P_{\rm neq}$ and
$\delta P_{\rm neq}^{(1)}$ computed as the functions of separation for
$\mu=0.25$~eV and the same temperatures $T_p^{(2)}=500$~K and $T_p^{(2)}=77$~K.

The short-dashed lines belonging to the pairs 1 and 2 in Fig. 1 (right)
behave similarly. Based on this, one can conclude that the nontrivial
behaviors of the solid lines representing $\delta P_{\rm neq}$ at nonzero $\mu$
is determined by the proper nonequilibrium contribution $\Delta P_{\rm neq}$
to the total nonequilibrium Casimir pressure in (\ref{eq12}).

Now we consider the dependence of the total nonequilibrium Casimir pressure
on the chemical potential at different separations. The energy gap parameter
is the same as above, i.e., $\Delta=0.3$~eV. In Fig. 2 (left), the
computational results for $P_{\rm neq}$ and $P_{\rm neq}^l$ are shown as the
functions of $\mu$ by the solid and long-dashed lines plotted at
$T_p^{(2)}=77$~K and $T_p^{(2)}=500$~K for the plates distant $a=0.2~\muup$m
apart. For comparison purposes, the solid and long-dashed lines plotted at
$T_p^{(2)}=T_E=300$~K illustrate the case of the equilibrium Casimir pressure
between the first and second plates computed using the full and the
spatially local theoretical approaches, respectively.
\begin{figure}[t]
\vspace*{-10.cm}
\hspace*{-6.mm}
\centerline{\includegraphics[width=17.5cm]{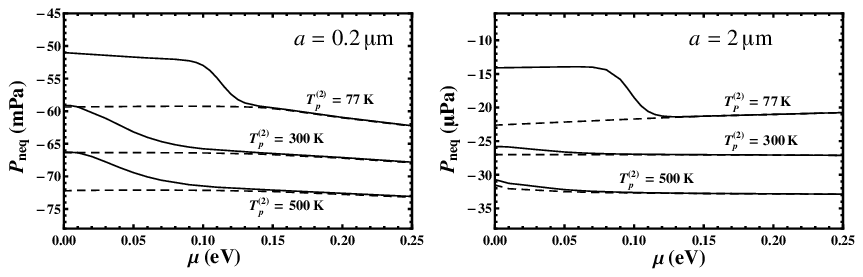}}
\vspace*{-11.3cm}
\caption{The computational results for the nonequilibrium Casimir
pressure, $P_{\rm neq}$, at $T_p^{(2)}=77$~K and 500~K in comparison with the
equilibrium one, $P_{\rm eq}$, at $T_p^{(2)}=300$~K are shown as the functions
of the chemical potential by the pairs of solid and dashed lines computed
within the spatially nonlocal and local theories, respectively, (left) at
$a=0.2~\muup$m and (right) at $a=2~\muup$m.
\label{f2}}
\end{figure}

{}From Fig. 2 (left), it is again seen that for all values of $\mu$ the
magnitude of the nonequilibrium Casimir pressure on a heated plate is
larger and on a cooled plate is smaller than the magnitude of the Casimir
pressure in the state of thermal equilibrium. From this figure it is also
confirmed that the contribution of the effects of spatial nonlocality
into the Casimir pressure increases with decreasing temperature.
An important new result is that with increasing chemical potential the role
of the effects of spatial nonlocality decreases and becomes negligibly
small at $\mu$ exceeding 0.2~eV.

In Fig. 2 (right), similar results are presented for the graphene-coated
plates spaced at $a=2~\muup$m. It is seen that with increasing separation
the role of the spatial nonlocality is suppressed at both $T_p^{(2)}=500$~K
and 300~K. However, for a graphene-coated plate cooled to $T_p^{(2)}=77$~K,
the spatial nonlocality plays the same role as at the shorter separation
distances within approximately the same range of $\mu$ and should be
taken into account in computations of the nonequilibrium Casimir pressure.

\section{Conclusions and Discussion}

In the foregoing, we have investigated the nonequilibrium Casimir pressure
on either the warmer or cooler graphene-coated plate than the environment
under an assumption that another plate is kept at the environmental temperature.
Calculations were performed in the framework of the Dirac model of graphene
using the polarization tensor, which takes into account the effects of
spatial nonlocality, as well as the energy gap and the chemical potential
of the graphene coating.

According to our results, the magnitude of the nonequilibrium Casimir pressure
on a warmer and cooler graphene-coated plate is larger and smaller than the
magnitude of the equilibrium Casimir pressure, respectively. Special attention
was paid to the role of the effects of spatial nonlocality. It is shown that
the spatially local theory underestimates the role of the effects of thermal
nonequilibrium. The contribution of the spatially nonlocal effects to the
nonequilibrium Casimir pressure turned out to be larger for a smaller
chemical potential of the graphene coating and for a cooler plate than the
environment.

The obtained results can be useful for the rapidly progressing field of
microelectronics exploiting graphene and other novel two-dimensional
materials.

\section*{Acknowledgments}

The work of G.L.K.and V.M.M. was supported by the
State Assignment for basic research (project FSEG--2023--0016).

\end{document}